# Understanding the Impact of Open-Framework Conglomerates on Water-Oil Displacements: Victor Interval of the Ivishak Reservoir, Prudhoe Bay Field, Alaska

Naum I. Gershenzon, Mohamadreza Soltanian, Robert W. Ritzi Jr., David F. Dominic

Department of Earth and Environmental Sciences, Wright State University, 3640 Col. Glenn Hwy., Dayton, OH 45435

Naum I. Gershenzon (naum.gershenzon@wright.edu)

**ABSTRACT**

The Victor Unit of the Ivishak Formation in the Prudhoe Bay Oilfield is characterized by high net-to-gross fluvial sandstones and conglomerates. The highest permeability is found within sets of cross-strata of open-framework conglomerate (OFC). They are preserved within unit bar deposits and assemblages of unit bar deposits within compound (braid) bar deposits. They are thief zones limiting enhanced oil recovery. We incorporate recent research that has quantified important attributes of their sedimentary architecture within preserved deposits. We use high-resolution models to demonstrate the fundamental aspects of their control on oil production rate, water breakthrough time, and spatial and temporal distribution of residual oil saturation. We found that when the pressure gradient is oriented perpendicular to the paleoflow direction, the total oil production and the water breakthrough time are larger, and remaining oil saturation is smaller, than when it is oriented parallel to paleoflow. The pressure difference between production and injection wells does not affect sweep efficiency, although the spatial distribution of oil remaining in the reservoir critically depends on this value. Oil sweep efficiency decreases slightly with increase in the proportion of OFC cross-strata. Whether or not clusters of connected OFC span the domain does not visibly affect sweep efficiency.

# INTRODUCTION

Prudhoe Bay Field, Alaska, is the largest oil field in North America, and has been under production since 1977 (Morgridge and Smith, 1972; Jones and Speers, 1976; Jamison et al., 1980). Oil, condensate, and gas are produced from the Triassic Ivishak Formation (Figure 1) occurring on the south limb of the Barrow Arch (an east-west trending anticline). Though large in total area and total reserves, the reservoir is compartmentalized by bounding faults into structurally isolated subunits on the order of 100 acres or smaller (Tye et al., 2003). The Ivishak Formation has a fluvial origin (Jones and Spears, 1976; Wadman et al., 1979; Atkinson et al., 1990; Tye et al, 1999) with a very high net-to-gross (0.73 to 0.96). As a result, the fluvial architecture can play a significant role in reservoir depletion under both gravity drainage and enhanced oil recovery (EOR, including waterflooding, miscible-gas flooding, and gas cycling). The understanding of this structure and this sedimentary architecture has aided the customization of well designs and completion types in structurally isolated and heterogeneous targets immediately below the gas cap (Tye et al., 2003).

The Victor Unit of the Ivishak Formation (Figure 1) is one of the more important units for production. It comprises fluvial sandstones and conglomerates, with sparse and patchy mudstones. Characteristically it is 46 m thick, has a net to gross of 0.96, and has permeability that ranges from $10^{-1}$ D to $10^2$ D. The highest permeability is found within sets of cross-strata of open-framework conglomerate (Figure 2). These cross strata are deposited through accretion of river gravels on the lee side of migrating subaqueous dunes under conditions that prevent sand accumulation and create grain-supported textures (Lunt and Bridge, 2007). They are preserved within unit bar deposits and the larger assemblage of unit bar deposits within compound (braid) bar deposits as shown in Figure 3 (Lunt et al., 2004). Though not originally recognized as open-framework conglomerates, these well-connected zones of high permeability were long known, as was their influence as thief zones in EOR (Atkinson et al., 1990, McGuire et al., 1994). These high-permeability zones have been represented in reservoir simulation studies analyzing the performance of EOR strategies (Stalkup and Crane, 1994; McGuire and Stalkup, 1995; McGuire et al., 1995, 1998). Important to the discussion below, the simulations developed in these previous studies were two-dimensional, and also predated extensive research on open-framework gravels and conglomerates (e.g. Lunt et al., 2004a). Furthermore, these simulations did not represent how capillary pressure relationships vary among different lithofacies. In this article we present three-dimensional simulations that include new information about the sedimentary architecture of open-framework conglomerates, and we explore the importance of capillary pressure relationships to residual oil saturation.

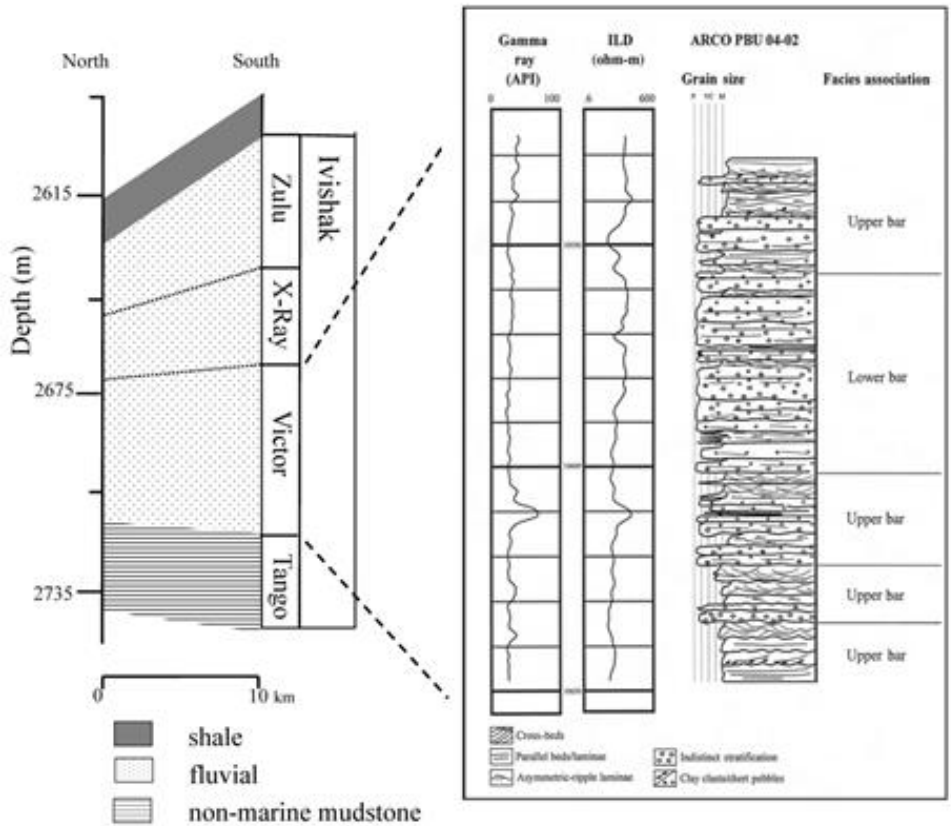

**Figure 1.** Prudhoe Bay stratigraphic information. (Left) Generalized stratigraphic section emphasizing producing zones within the Ivishak Fm. including the Victor Interval, which is the focus of this paper (adapted from Tye et al., 1999). (Right) Typical logs (gamma ray and induction) for the Victor Interval (from Tye et al., 2003). The Victor interval is dominated by fluvial, compound-bar deposits.

The importance of open-framework conglomerates has motivated extensive research clarifying aspects of their formative fluvial processes, and quantifying important attributes of their sedimentary architecture within preserved deposits (Tye et al., 2003; Lunt et al., 2004a, 2004b, 2007; Bridge 2006). This work has led to new quantitative facies models for fluvial deposits in general, including those comprising open-framework conglomerates. Particularly relevant are synthesized studies of the Sagavanirktok River on the Alaska North Slope, a modern analog for the depositional setting of the Ivishak Formation. Cross strata of open-framework gravel are decimeters thick and meters to tens of meters in lateral extent (Lunt et al., 2004). They are commonly bounded by strata of bimodal sandy gravel (Figure 4). These cross strata occur within large-scale inclined strata associated with unit bar migration which occur, in turn, within still larger scale strata associated with compound bar accretion. In these compound-bar deposits, open-framework gravels were found with volume fractions on the order of 0.3 (Lunt et al., 2004a,b; Bridge 2006).

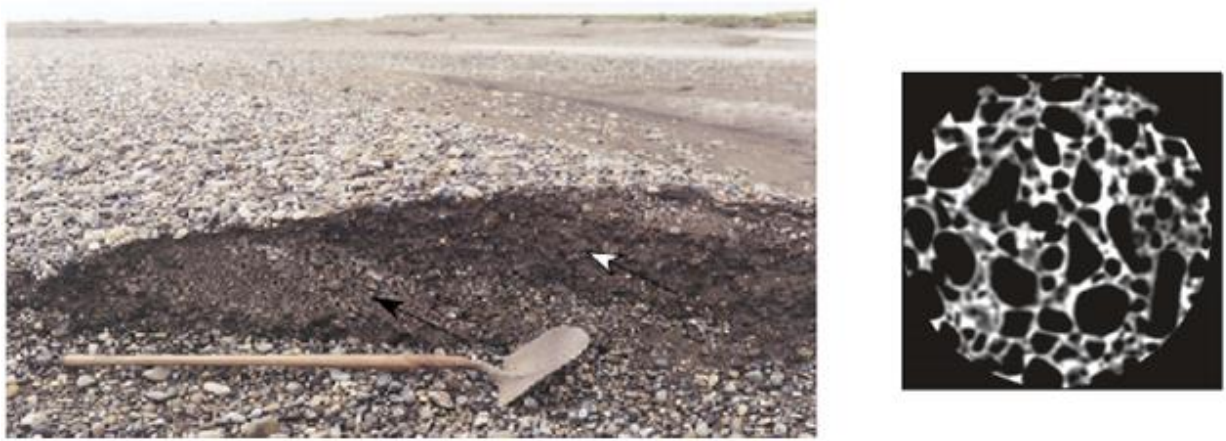

**Figure 2.** (Left) Dissected unit bar showing alternations of large-scale inclined sandy gravel (white arrow) and open-framework gravel (black arrow) cross strata. Shovel is 1.3 m long (from Lunt, 2002). (Right) Axial CT-scan slice through a cored open-framework conglomerate sample from the Victor Interval of the Ivishak Fm (from Tye et al., 2003). The sample is 4 inches wide. The black areas are chert pebbles. The white area is pore space.

The saturated permeability in sandy-gravel deposits varies non-linearly as a function of the volume of sand mixed with gravel (see Figure 6 in Ramanathan et al., 2010; and also Klingbeil et al., 1999; Conrad et al., 2008; and Porter et al., 2012). Sandy gravel strata have permeabilities similar to the sand they contain, which are of the order of $10^0$ to $10^1$ D. Thus, sand and sandy gravel strata within unit bars have permeabilities similar to channel-fill sands. Open-framework gravels have permeabilities of the order of $10^3$ to $10^4$ D (Klingbeil et al., 1999; Ferreira et al., 2010). In either type of strata, the coefficient of variation in permeability is of the order of unity. In the lithified stratatypes within the Ivishak Formation (sandstones, pebbly sandstones and open-framework conglomerates), the saturated permeabilities scale down accordingly (Tye et al., 2003).

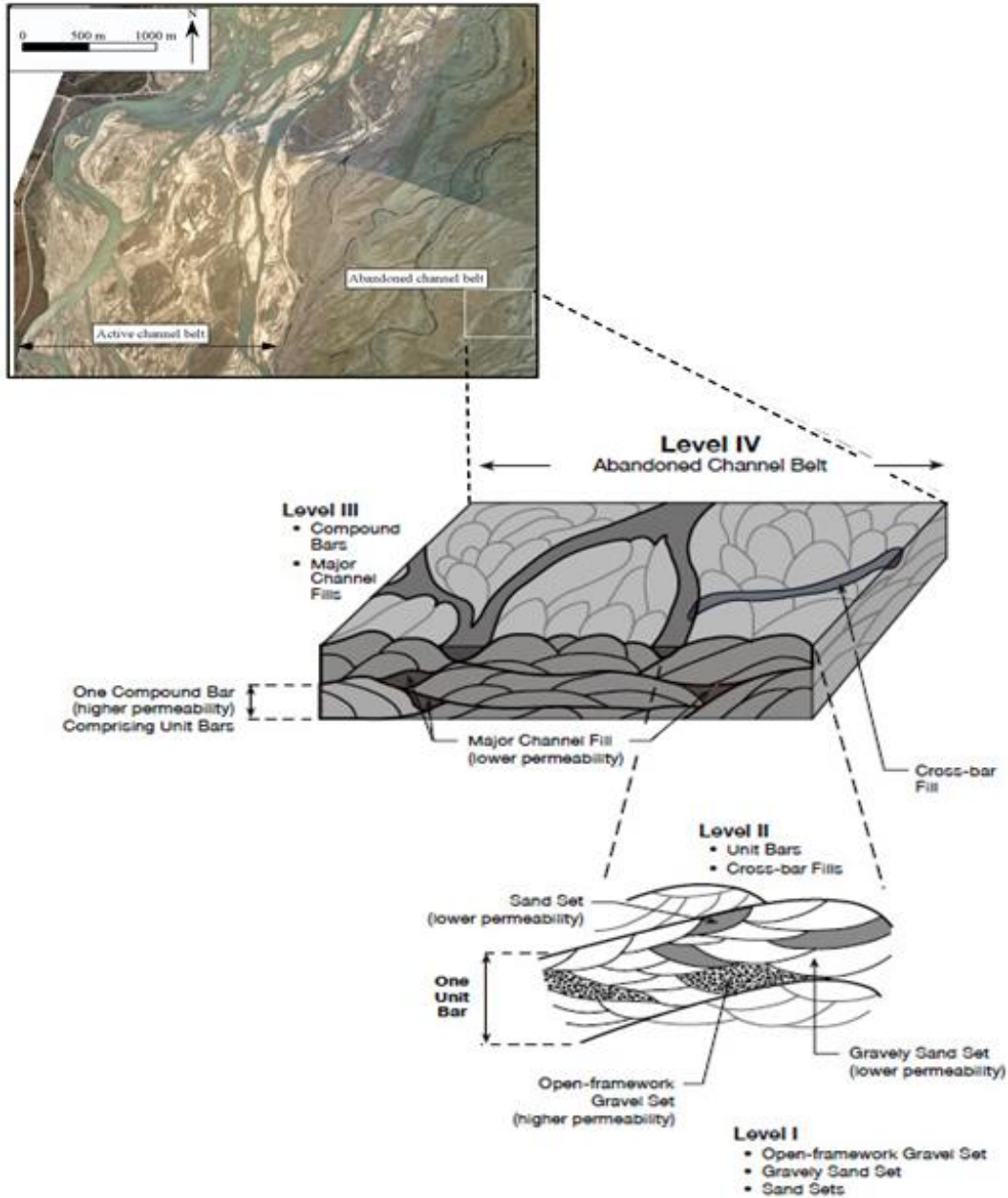

**Figure 3.** (Top) Study area in the active channel belt and in the preserved channel belt deposits of the Sagavanirktok River (Lunt et al. 2004). (Middle and Bottom) Conceptual model for the hierarchical sedimentary architecture found in channel belt deposits (see also Table 1). The compound bar deposits at level III result from the processes of unit-bar accretion and channel migration. Within unit bar deposits (level II), cross-strata of open-framework gravel (level I) have highest permeability. As channels are abandoned, they are filled with lower-permeability sediment. Major channel fills (level III) and smaller cross-bar channel fills (level II) are lower-permeability baffles within the deposit. From Ramanathan et al. (2010).

**Table 1.** Hierarchy of Unit Types

| IV | channel-belt deposit | | | | |
|---|---|---|---|---|---|
| III | compound bar deposits[1] | | | | major channel fills |
| II | unit bar deposits | | | cross-bar channel fills | concave-up sand |
| I | open-framework gravel set[2] | gravelly sand set | sand set | concave-up sand | concave-up sand |

[1.] typical dimensions (largest unit type): *750 x 500 x 2 m³*

[2.] typical dimensions (smallest type): decimeters to meters long and wide, centimeters to decimeters thick

    The saturated permeability in sandy-gravel deposits varies non-linearly as a function of the volume of sand mixed with gravel (see Figure 6 in Ramanathan et al., 2010; and also Klingbeil et al., 1999; Conrad et al., 2008; and Porter et al., 2012). Sandy gravel strata have permeabilities similar to the sand they contain, which are of the order of $10^0$ to $10^1$ D. Thus, sand and sandy gravel strata within unit bars have permeabilities similar to channel-fill sands. Open-framework gravels have permeabilities of the order of $10^3$ to $10^4$ D (Klingbeil et al., 1999; Ferreira et al., 2010). In either type of strata, the coefficient of variation in permeability is of the order of unity. In the lithified stratatypes within the Ivishak Formation (sandstones, pebbly sandstones and open-framework conglomerates), the saturated permeabilities scale down accordingly (Tye et al., 2003).

    The newer information on fluvial architecture was incorporated into a geocellular model by Ramanathan et al. (2010) and Guin et al. (2010). Relevant aspects of this model include: 1) It simulates many scales of hierarchical sedimentary architecture ranging from decimeter-scale strata to a kilometer–scale channel belt, and includes deposits of unit bars, compound bars, cross-bar channel fills and major-channel fills. The organization and resulting connectedness of open-framework conglomerates are influenced by all these scales. 2) A continuous-in-space stochastic-geometric modeling approach was used to facilitate an appropriate representation of the characteristic shape of these units, their variability in size, and their juxtapositioning relationships (e.g. conformable vs. erosional boundaries). 3) Sample statistics for the proportion and size of all unit types in the model were shown to honor the field-measured values of Lunt et

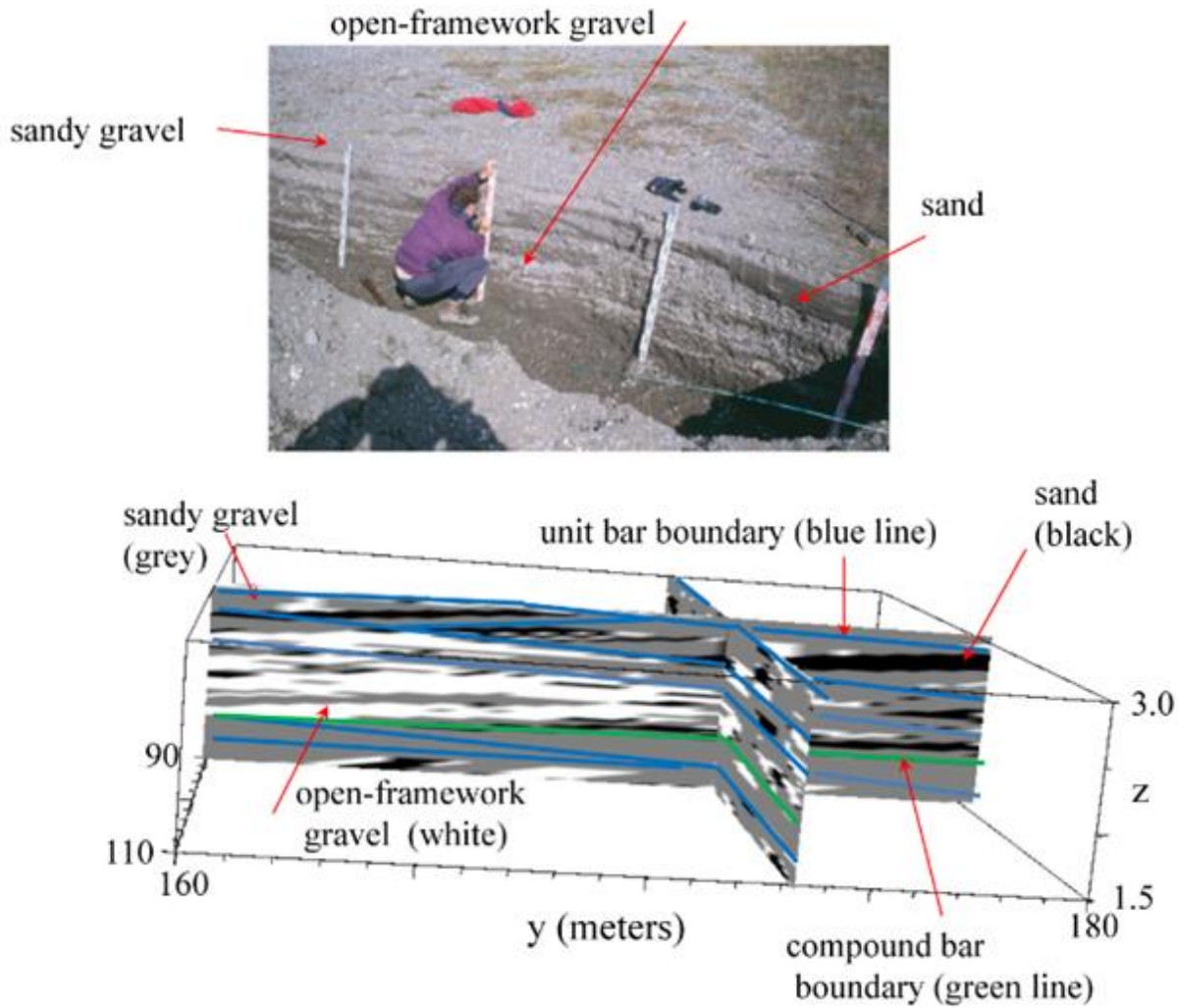

**Figure 4.** (Top) Exposure of level I unit types in a trench at the Sagavanirktok River field site (from Lunt, 2000). (Bottom) Rendering of orthogonal sections through an extracted piece of the stratal model produced for realization 1 with the GEOSIM code. The extracted piece was chosen so that open-framework gravel/conglomerate (28% of overall model volume) was clearly visible. Paleoflow direction is to the left.

al. (2004). 4) The continuous-in-space geometric model can be sampled with any desired grid resolution in creating a geocellular model. This allows permeability to be mapped into the geocellular model according to the lithofacies boundaries and at a finer resolution than these.. Figure 4 shows a cross section through an extracted piece of a simulated compound bar deposit sampled with fine resolution (larger scale simulations showing larger-scale architecture are given in Ramanathan et al. (2010, Figure 10). 5) The connectivity of open-framework conglomerates created in this model has been quantified and rigorously studied. To briefly explain, the literature from the branch of mathematics called Percolation Theory shows that randomly placed cells of one type (e.g. open-framework conglomerate) in the geocellular model grid will connect and form pathways that span opposing boundaries (along tortuous paths) if included with proportions above 0.31 (Stauffer and Aharony, 1994; Hunt and Ewing, 2009). Furthermore, the literature shows that such spanning pathways will form at even lower proportions in the presence of geologic spatial organization (Harter, 2005; Guin and Ritzi, 2008). Indeed, the cells of open-framework conglomerate in geocellular models form connected spanning pathways if simulated with proportion above 0.2 (Guin et al., 2010).

To date, this new information and new model have not been incorporated into reservoir simulations. Thus, there is an opportunity to newly examine the influence of open-framework conglomerates on multi-phase flow and on reservoir performance by running relatively high-resolution simulations using this stratal model. Note that the stratal model is not based explicitly on data from the Victor interval but is thought to represent important aspects of it. and thus the simulation model is designed to be generic for the reservoir. The simulations we present are screening studies which investigate the impact of this type of sedimentological heterogeneity within reservoir layers.

The complexity of the heterogeneity and highly non-linear nature of the equations representing flow make it challenging to achieve numerically convergent solutions. Consequently, the simulations presented here are limited to an examination of water flooding [e.g. Ringrose et al. 1993; Gharbi et al, 1997; Choi et al, 2011] with the black oil approximation, and in a relatively small reservoir compartment (though our longer-term goal is to run larger compositional simulations of miscible gas injection). Sweep efficiency in a waterflood is fundamentally controlled by the nature of immiscible displacement of non-wetting liquid by wetting liquid in porous media (Buckley and Leverett, 1942), which includes the effects of capillary pressure and relative permeability on oil trapping and water breakthrough (Kortekaas, 1985; Corbett et al, 1992; Khataniar and Peters, 1992; Wu et al, 1993; Gharbi et al, 1997; Kaasschieter, 1999). Such immiscible displacement is, in turn, controlled by the three-dimensional architecture of reservoirs (Kjonsvik et al., 1994; Jones et al, 1995; Tye et al, 2003; Choi et al, 2011). We expect that differences in capillary pressure relationships between open-framework conglomerates and other lithofacies, as well as differences in relative permeability, are important to understanding sweep efficiency. Furthermore, we expect that connectivity is important. By using models for which the connectivity of each cell of open-framework conglomerate is known and quantified, we can segregate, quantify, and study the different roles that connected and unconnected open-framework conglomerates have on immiscible displacement processes.

**METHODOLOGY**

We simulated three-dimensional immiscible oil displacement by water (black oil approximation) with ECLIPSE (Schlumberger Reservoir simulation software, version 2010.2). The approach was as follows. The top of the parallelepiped reservoir model is taken to be at a depth of 2560 m, which is equivalent to the top of the Ivishak Formation. The size of the reservoir in the *x*, *y* and *z* directions are $L_x$= 200 m, $L_y$= 200 m and $L_z$= 5 m, respectively. Initially the reservoir contains oil with dissolved gas and connate water in equilibrium. The reservoir is divided into one million cells with cell size 2 m x 2 m x 0.05 m in the *x*, *y* and *z* directions, respectively.

Using a code by Ramanathan et al. (2010) we generated six realizations of the channel-belt architecture, each with a different proportion of open-framework conglomerate (Table 2). From each realization, we extracted a sample block having the size of the reservoir model. Each block came from within compound bar deposits. As stated above, sandy-gravel strata have permeabilities similar to the sand they contain. In this study, the geometric mean for saturated permeability of both sandstone and sandy conglomerate is taken to be 66 mD. Because these two lithotypes have the same permeability distribution in the simulations, we refer to them collectively as "sandstone." The geometric mean saturated permeability of open-framework conglomerates is taken to be 5250 mD. For each lithotype, saturated permeability distributions were created with a coefficient of variation of unity as is found in natural deposits. The permeability of each cell is assigned from the appropriate distribution as of the lithotype it contains.

As shown in Table 2, the proportion of open-framework conglomerates (hereinafter simplified as OFC) in the samples control the mean permeability, and also affects the proportion of open-framework conglomerate cells that are connected to create preferential flow pathways. Clusters of OFC cells "span" opposing boundaries (along tortuous paths) when their extent in *x*, *y* and *z* directions is equal to the reservoir size in those directions. Thus, the reservoir models range from one with more than 90% of all OFC cells connected in one spanning cluster to those with no spanning cluster. Spanning occurs at proportions above 20%, consistent with results by Guin et al. (2010).

**Table 2**

| Realization | OFG Proportion (%) | Geometric Mean Permeability (mD) | Proportion of Connected OFG Cells Among All Cells (%) | Proportion of Connected OFG Cells Among OFG Cells (%) | Do Clusters Span Opposing Boundaries? |
|---|---|---|---|---|---|
| 1 | 28 | 226 | 26 | 91 | yes |
| 2 | 26 | 205 | 22 | 85 | yes |
| 3 | 24 | 193 | 17 | 71 | yes |
| 4 | 22 | 174 | 12 | 53 | yes |
| 5 | 19 | 149 | 1.3 | 6.8 | no |
| 6 | 16 | 136 | 0.15 | 0.9 | no |

Two different property tables were utilized for sandstone and OFG lithotypes. These tables define the relations between water relative permeability, oil-in-water relative permeability, water-oil capillary pressure, gas relative permeability, and water saturation. Since there is no reliably measured relative permeability for sandstone and OFC, we used the conventional approach (e.g., Jensen and Falta, 2005) to define the property tables utilizing simple rules: (1) relative permeability curves are limited by irreducible wetting phase saturation and residual non-wetting-phase saturation; (2) the maximum relative permeability for the wetting phase is less than the maximum relative permeability for the non-wetting phase; (3) the relative permeability for the wetting phase is less than the relative permeability for the non-wetting phase for the same saturation of each phase; (4) the total permeability of two phases of immiscible liquid is always less than the permeability of one phase. The relative permeability for oil ($k_{ro}$) and water ($k_{rw}$) have been calculated by the formulae used by Kortekaas (1985):

$$k_{ro} = 0.85(1 - S_e)^3, k_{rw} = 0.3(S_e)^3, S_e = \frac{S_w - S_{wir}}{1 - S_{or} - S_{wir}}$$

where $S_w$ is the saturation of the wetting phase (water in our case), $S_{wir}$ is the irreducible saturation of the wetting phase, $S_{or}$ is residual oil saturation. Figure 5 shows the relative permeability as a function of water saturation for sandstone and OFC lithotypes. We used the Brook and Corry (1964) approach to characterize capillary pressure, $P_c$ (psi):

$$P_c = 14.5 C S_e^{-\frac{1}{\lambda}} (\frac{\phi}{k})^\alpha ,$$

where φ is the porosity (φ=0.2 for all calculations, Tye et al. (2003)); $k$ is the permeability in mD; and α, λ and C are constants with these values: α=0.5, and λ=1.5, $C$ = 5. Figure 6 shows the results.

At the reference pressure of 4014.7 psi (hydrostatic pressure at depth 2560 m), the water formation volume factor is 1.029 rb/stb = rm$^3$/sm$^3$, the water viscosity is 0.31cP, and the water and rock compressibility are 3.13·10$^{-6}$ 1/psi and 3.13·10$^{-6}$ 1/psi, respectively. The oil, water, and gas gravities at surface conditions are 23.5, 1.04, and 0.8, respectively.

In the simulations, we utilized values for the pressure difference between injection and production wells from 100 to 800 psi (0.689 to 5.52 MPa). For simplicity, pressure difference was kept constant during a single simulation and pressure inside the reservoir was always kept above the bubble-point. Injection and production wells were vertical; in some simulations these were aligned in the *y* direction (i.e., parallel to paleoflow) and in others they were aligned in the *x* direction (i.e., perpendicular to paleoflow).

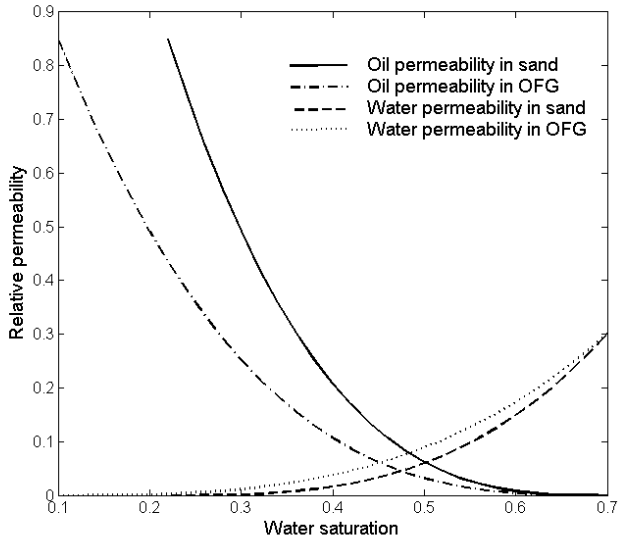

**Figure 5.** Relative permeability of oil and water versus water saturation for sand and for open-framework gravel.

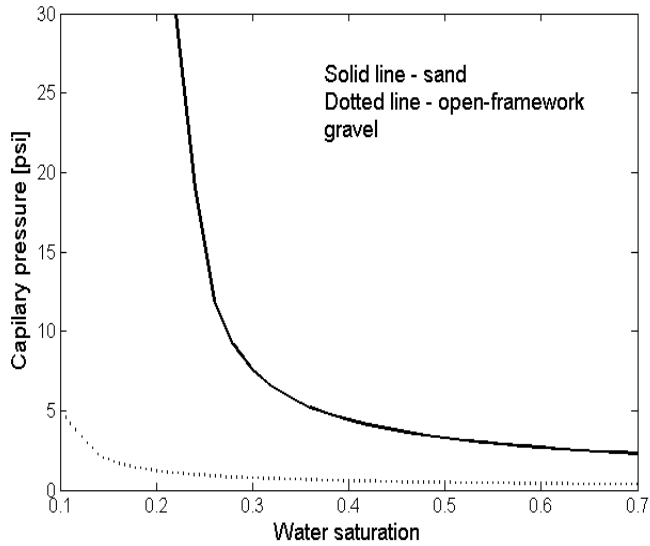

**Figure 6.** Capillary pressure versus water saturation for sand (solid line) and for open-framework gravel (dotted line).

1. **RESULTS**

**Anisotropy**

Fluvial deposits are expected to be anisotropic with different permeability and mobility (ratio of permeability to viscosity) in the paleoflow direction ($y$), in the direction normal to the paleoflow ($x$), as well as in the vertical direction ($z$). To consider how anisotropy affects the process of oil displacement, we compared results of simulations with the pressure gradient along the $y$ direction to the same case with the pressure gradient along the $x$ direction. Figures 7 and 8 are maps of oil saturation obtained from the simulations. Figure 8 conveys the effects of anisotropy. Under the same pressure gradient between injection and production wells, the waterflood front propagates faster when the pressure gradient is in the $y$ direction. Importantly, Figure 8 shows that the waterflood front is broader when the pressure gradient is in the $x$ direction, so that a larger volume of the reservoir is swept. This qualitative result indicates that sweep efficiency increases if the injector and producer are aligned perpendicular to the paleoflow direction, a result that is quantified below.

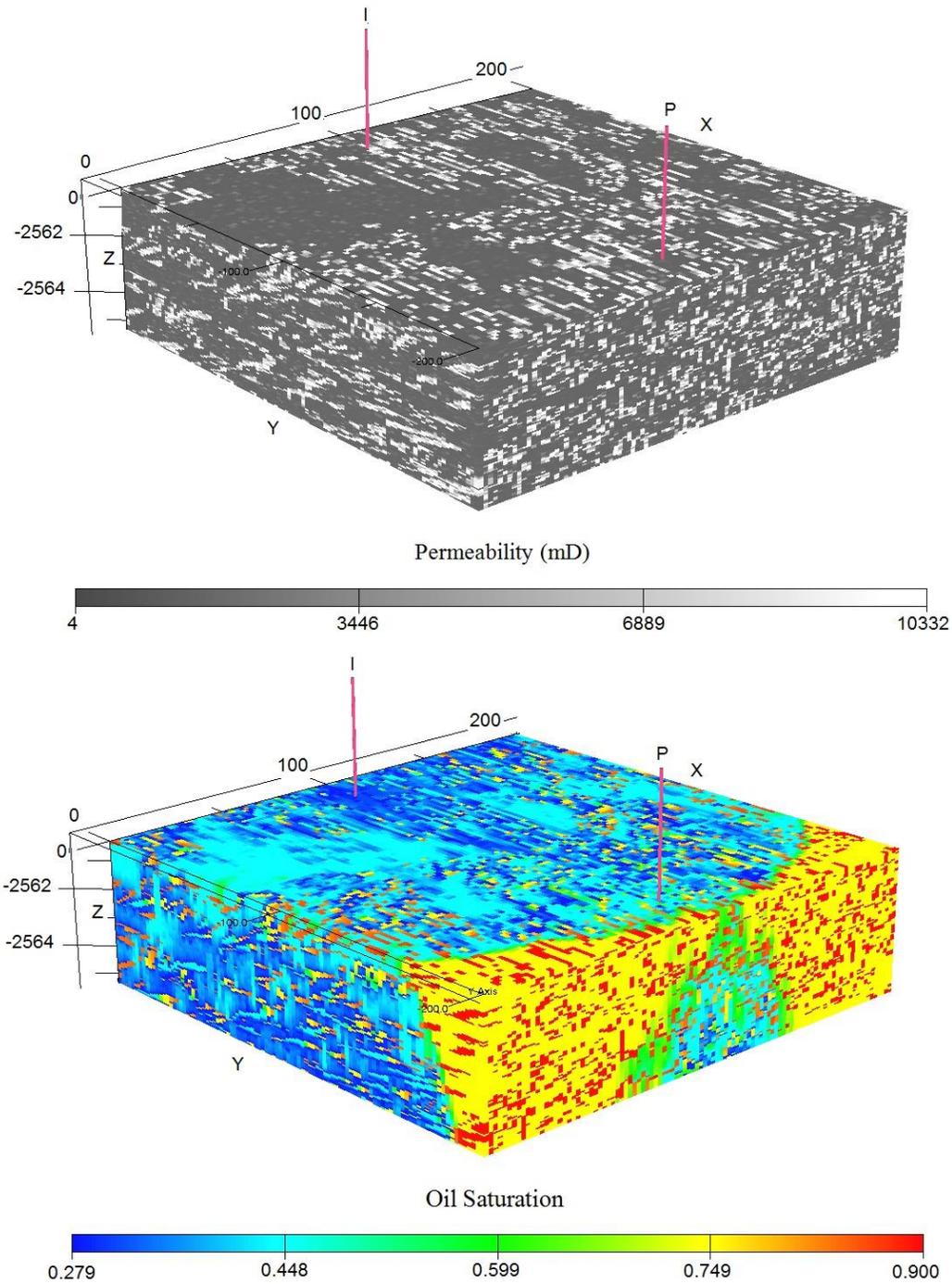

**Figure 7.** (Top) Permeability, realization #3. (Bottom) Oil saturation after 90 days when injection well (I), production well (P), and pressure gradient are aligned along the y direction. Pressure difference between wells is 200 psi. In the regions beyond the waterflood front, the cells with highest initial saturation are open-framework gravel. In the regions behind the waterflood front, those cells with highest residual saturation are also open-framework gravel.

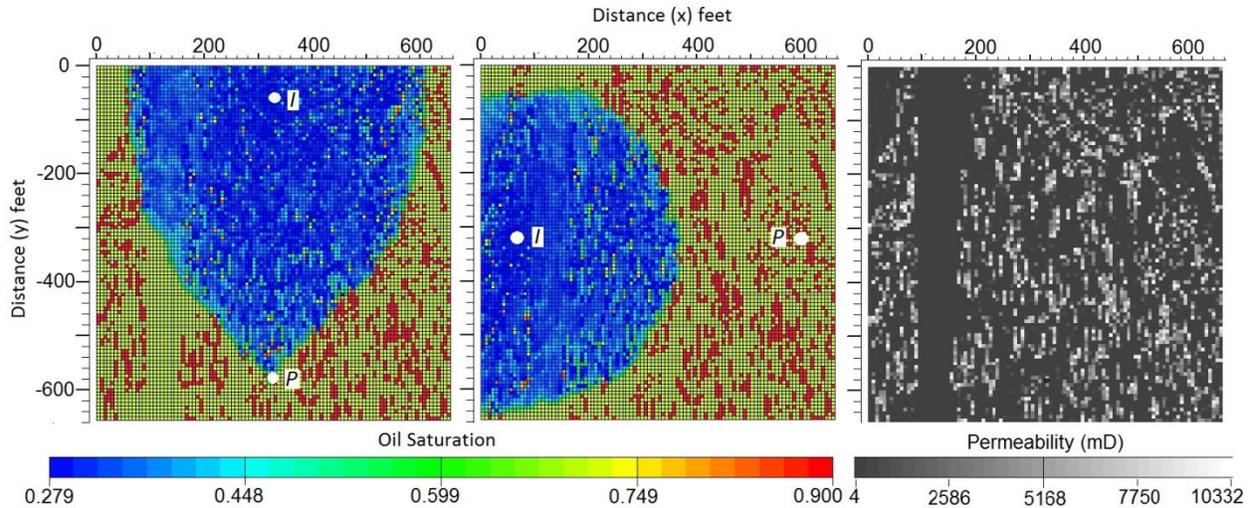

**Figure 8**. Horizontal sections along the middle layer of realization #3 showing oil saturation after 50 days. (Left) Injection well (I), production well (P), and pressure gradient are aligned along the *y* direction: (Middle) alignment along the *x* direction. Note front of the waterflood is broader and migrating slower; (Right) Permeability map. Pressure difference between wells is 200 psi.

When the pressure difference between injection and production wells is constant, production rate can be used as a measure of effective mobility. This is the case for Figure 9, which shows oil and water production rates versus injected water volume for each of the modeled realizations. Figure 9 illustrates that the dynamics of oil and water production change progressively among realizations for which the proportion of OFC decreases from 28% to 16%. These results show that mobility in the *y* direction is greater than in the *x* direction. Indeed the breakthrough time in the latter case ranges from 1.35 to 1.44 times larger than in the former case for all realizations. The higher mobility in the *y* direction decreases sweep efficiency.

Figure 9 also shows that when the pressure gradient is in the *y* direction, the production rate is initially larger than if the gradient is in the *x* direction, for all realizations except #6. However, water breakthrough is earlier in this case for all realizations. This earlier water breakthrough results in lower oil sweep efficiency. Figure 10 shows that this effect is reflected in the oil saturation averaged over the entire reservoir both at the time of water breakthrough and after the injection of one effective movable pore volume (pore volume minus volume of connate water and volume of irreducible oil), and is practically independent of the pressure difference.

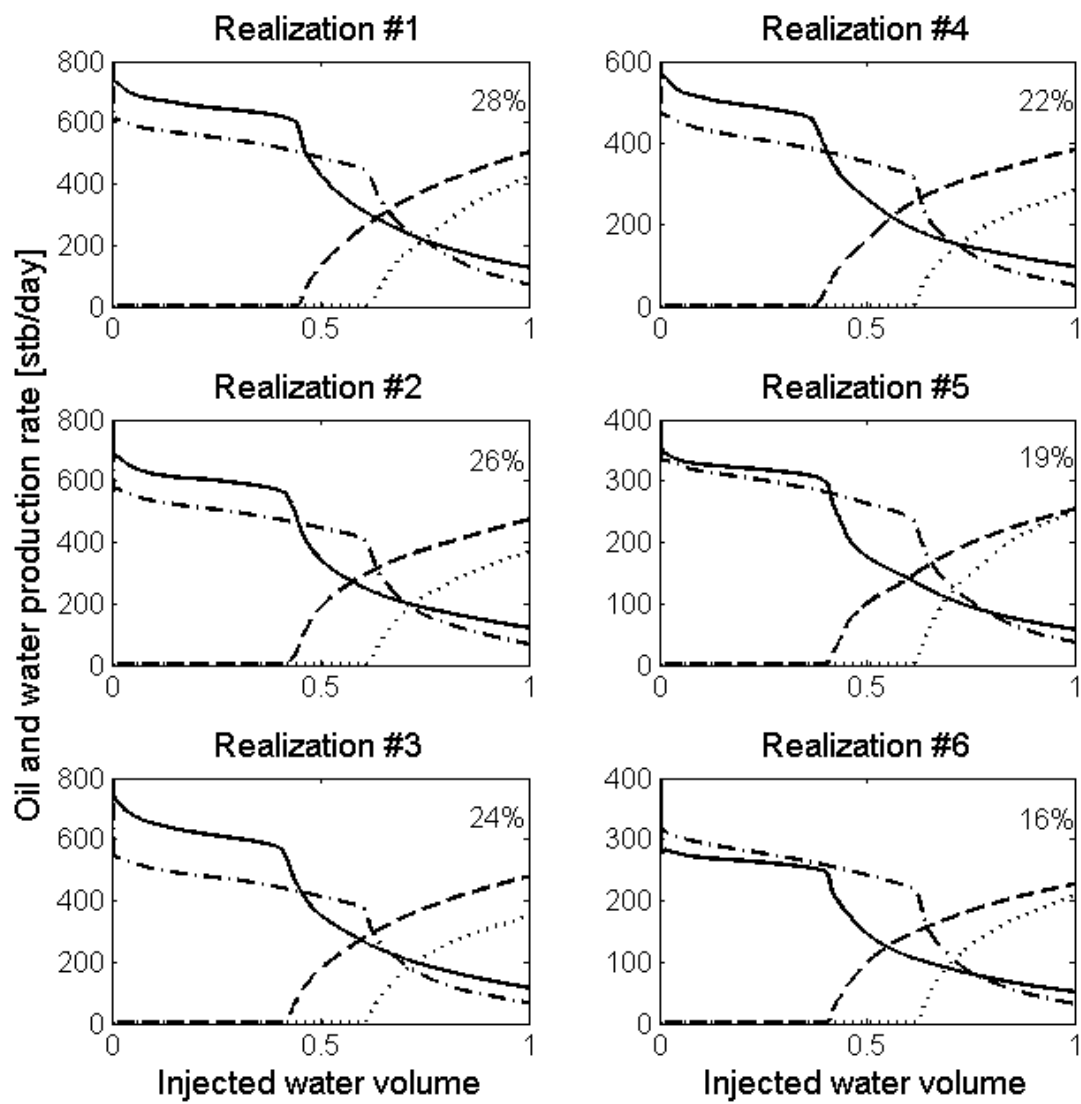

**Figure 9**. Oil and water production rates versus injected water volume for the six realizations (proportion of OFG is shown within each graph). Pressure difference is 200 psi. Solid lines show oil production rates when the pressure gradient is along the *y* direction; dashed lines show water production rates when pressure gradient is along the *y* direction; dot-dashed lines show the oil production rates when pressure gradient is along the *x* direction; dotted lines show water production rates when pressure gradient is along the *x* direction. Injected water volume is normalized by the pore volume of the reservoir minus irreducible water volume and residual oil saturation

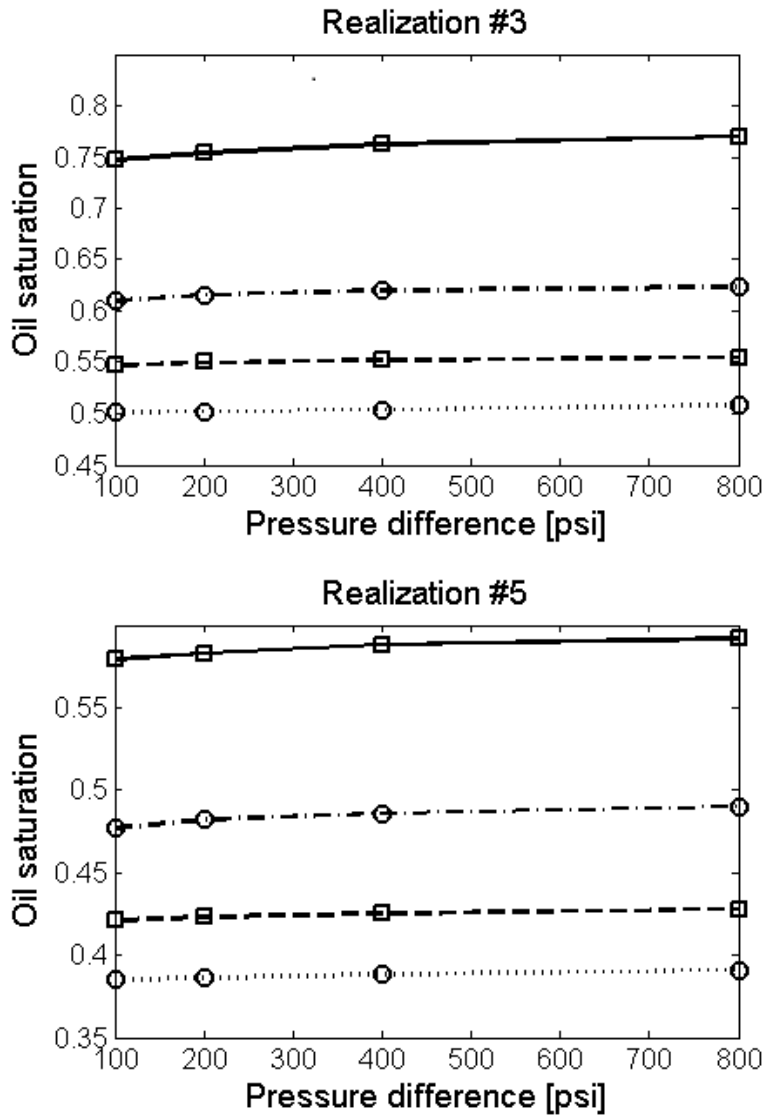

**Figure 10**. Oil and water production rates versus injected water volume for the six realizations (proportion of OFG is shown within each graph). Pressure difference is 200 psi. Solid lines show oil production rates when the pressure gradient is along the *y* direction; dashed lines show water production rates when pressure gradient is along the *y* direction; dot-dashed lines show the oil production rates when pressure gradient is along the *x* direction; dotted lines show water production rates when pressure gradient is along the *x* direction. Injected water volume is normalized by the pore volume of the reservoir minus irreducible water volume and residual oil saturation

Table 3 shows cumulative oil production as a proportion of movable pore volume for all realizations, up to water breakthrough and up to the injection of one movable pore volume of water. Results are given both for a pressure gradient along the *y* direction and along the *x* direction. Although oil production varies from realization to realization, it is always larger when the pressure gradient is along the *x* direction. For these realizations, cumulative oil production up to water breakthrough is about 50% larger on average when the gradient is along the *x* direction. We quantify sweep efficiency (*SE*) as the percentage of oil removed from a reservoir before water breakthrough: $SE = (S_{init} - S_{break})/S_{init}$, where $S_{init}$ and $S_{break}$ are initial and resulting (after water breakthrough) oil saturations. Using this formula and assuming that $S_{init} = 1$ we find the ratio between sweep efficiency when the pressure gradient is in the x direction and when it is in the y direction (see 4$^{th}$ column in Table 3).

**Table 3.** Total oil production normalized by movable pore volume. The pressure difference between wells is 200 psi.

| Realization | Up to water breakthrough (*y* direction) | Up to water breakthrough (*x* direction) | Sweep efficiency ratio between x and y directions (%) | Up to injection of one movable pore volume water (*y* direction) | Up to injection of one movable pore volume water (*x* direction) |
|---|---|---|---|---|---|
| 1 | 0.279 | 0.379 | 16.1 | 0.418 | 0.545 |
| 2 | 0.263 | 0.378 | 18.4 | 0.434 | 0.516 |
| 3 | 0.257 | 0.384 | 20.6 | 0.427 | 0.538 |
| 4 | 0.235 | 0.398 | 27.1 | 0.349 | 0.467 |
| 5 | 0.245 | 0.371 | 20.0 | 0.341 | 0.419 |
| 6 | 0.254 | 0.366 | 17.5 | 0.279 | 0.483 |

**Connectivity**

Surprisingly, Figure 9 shows that the production dynamics are not noticeably different between realizations with spanning OFC clusters (realizations 1-4) and those without spanning clusters (realizations 5 and 6). This observation is confirmed in Figure 11, which shows residual oil saturation averaged over the reservoir versus the proportion of OFC for each of the six realizations. This figure again illustrates the consistent difference that occurs with different orientations of the pressure gradient (circles versus squares; x versus diamonds), that these differences are practically independent of pressure difference (solid lines versus dashed lines), and that these differences persist from the time of water breakthrough (two upper lines) through the injection of one moveable pore volume of water (two lower lines). Although there is a weak trend of decreasing residual oil saturation as proportion of OFC increases, there is no clear difference in these results for realizations with spanning clusters (realizations 1 to 4) and those without spanning clusters (realizations 5 and 6). To evaluate stochastic uncertainty, simulations were further created with new and different seed numbers to generate realizations with proportions of OFC around 0.25. The results, also in Figure 11, show that differences in relative saturation between the *x* and *y* directions stand out above the stochastic uncertainty. Indeed, the scatter of relative oil saturation (at water breakthrough time) for different realizations is visibly smaller than differences in saturation between the *x* and *y* directions. The weak trends in residual saturation vs. proportion of OFC do not stand out above this scatter, supporting our contention that there are not clear differences as OFC proportion varies.

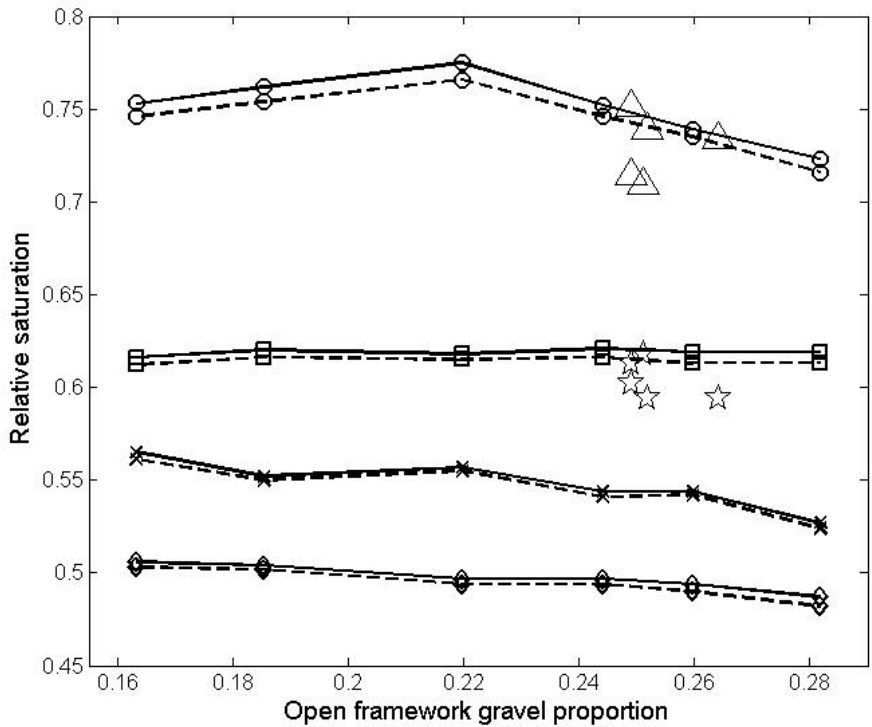

**Figure 11**. Averaged relative oil saturation versus OFG proportion in the six reservoir realizations. Pressure difference is 200 psi (dashed lines) and 400 psi (solid lines). Circles: oil saturation at water breakthrough, pressure gradient along the *y* direction; squares: oil saturation at water breakthrough, pressure gradient along the *x* direction; x: oil saturation after injection of one movable pore volume of water, pressure gradient along the *y* direction; diamonds: oil saturation after injection of one movable pore volume of water, pressure gradient along the *x* direction. The results of simulations (i.e. oil saturation at water breakthrough time at pressure difference 200 psi) with additional 5 realizations shown by stars and triangles, respectively for the pressure gradient in x- and y-directions.

Figure 11 illustrates results averaged over the entire reservoir for the different realizations, whereas Figures 12 and 13 illustrate results for individual cells. In these cases, clear differences can be seen. Figures 12 and 13 depict oil saturation in two nearby cells in the same realization. The cell depicted in Figure 12 is part of a large, spanning OFC cluster whereas the one in Figure 13 is part of a cluster that includes only 12 cells. For a cell within a spanning cluster, oil saturation diminishes rapidly as the waterflood front passes this cell, an effect that is practically independent of pressure difference. For a cell within a small cluster, oil saturation remains high and clearly depends on the pressure gradient. Figure 13 illustrates the effect, identified by Kortekaas (1985), that oil can be trapped in isolated heterogeneities, an effect that clearly depends on the pressure gradient, as previously identified by, for example, Corbett et al. (1992) and Ringrose et al. (1993).

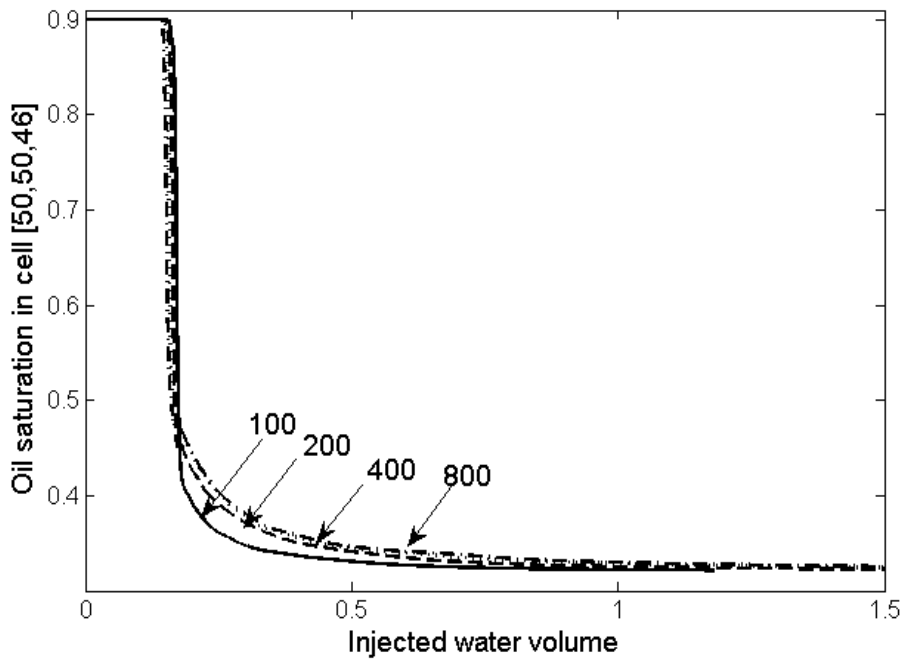

**Figure 12**. Oil saturation in realization 3 for a single cell of OFG as a function of injected water volume. The cell [50,50,46] is a part of a spanning cluster that includes 71% of all OFG cells in the realization. The results show little dependence on pressure difference over the range from 100 to 800 psi (labeled in the diagram).

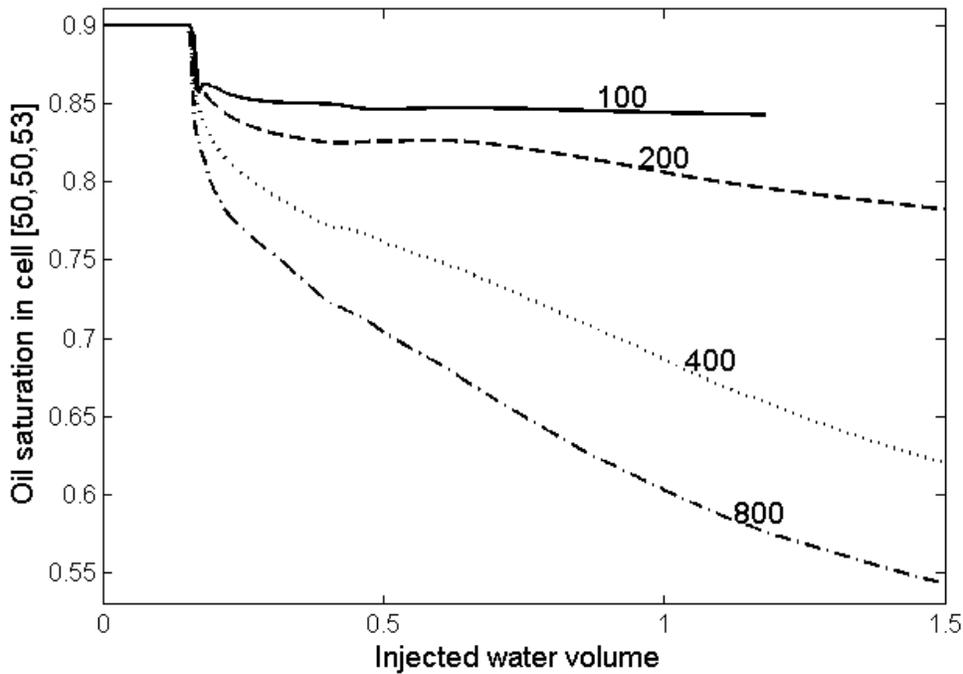

**Figure 13**. Oil saturation in realization 3 for a single cell of OFG as a function of injected water volume. The cell [50,50,53] is a part of a small cluster that includes only 12 connected OFG cells; the cluster extends 4 m in the x direction, 6 m in the y direction, and 0.25 m in the z direction. The result show considerable dependence on pressure difference over the range from 100 to 800 psi (labeled in the diagram).

Figure 14 shows oil saturation as a function of pressure difference for all OFC cells, all sandstone cells, and all cells within the model. As expected the oil trapping effect in OFC cells is clearly indicated by the higher overall values, with a clear dependence on the pressure difference. For OFC cells, oil saturation decreases as pressure difference increases. For sandstone, oil saturation is always much lower, but increases as pressure difference increases. In this case, capillary pressure pushes oil from sandstone cells to OFC cells. Importantly, oil saturation averaged over the whole reservoir is practically independent of the pressure difference, showing no net trapping effect. This unexpected result has been confirmed for all six realizations and occurs regardless of the orientation of the pressure gradient.

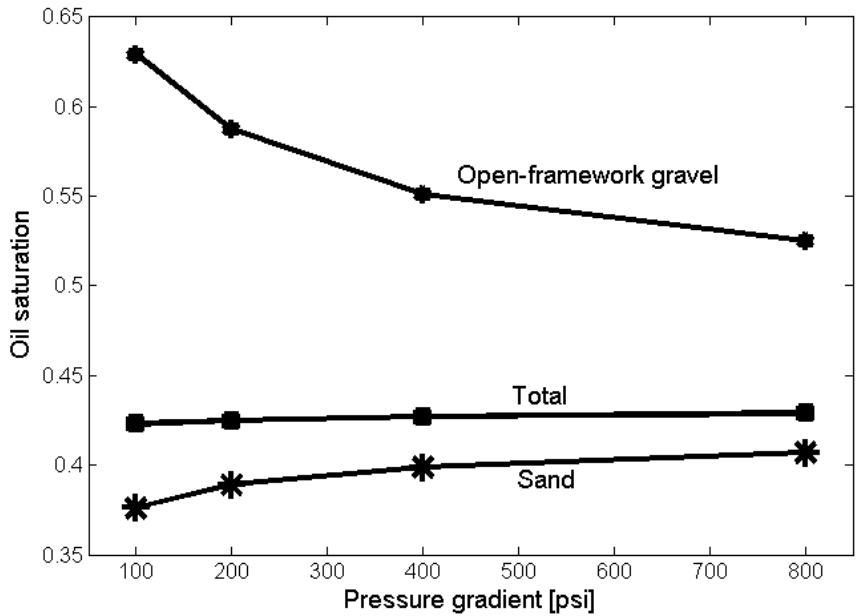

**Figure 14**. Oil saturation after injection of one movable pore volume of water versus pressure difference. Oil saturation is averaged over OFG cells, sand cells, and all cells. These results are for realization 5 with the gradient oriented along the *x* direction.

**DISCUSSION AND CONCLUSIONS**

Previous work has shown that during immiscible oil displacement, oil may be trapped when reservoir heterogeneities result from (1) uniform permeability but heterogeneous capillary pressure (Kortekaas, 1985) or (2) non-uniform permeability (Wu et al, 1993; Kaasschieter, 1999). The effect of capillary pressure depends on the pressure gradient ($\Delta P$) and the size of the heterogeneity in the direction of pressure gradient (*l*) (Corbett et al., 1992). If the difference in capillary pressure between two materials is small compared to $\Delta P*l$, the effect of capillary pressure in trapping oil is also small. In contrast, if the difference in capillary pressure is larger than $\Delta P*l$, the trapping effect can be considerable. As expected, our results show that more oil is trapped within the smallest OFC clusters (Figure 12) than in the large, spanning OFC clusters (Figure 13).

In our case, capillary pressure differs between sandstone (high capillary pressure) and OFC (low capillary pressure) such that residual oil saturation is always lower in sandstone. The interplay of processes controlled by the interconnection of OFC and those resulting from the juxtaposition of sandstone and OFC defines the spatial and temporal distribution of oil in our reservoir models. Given the heterogeneity within these models, it is surprising to find that oil saturation averaged over the reservoir does not depend on pressure difference. In essence, the effect of oil trapping in isolated OFC clusters is canceled by the effect of oil moving out of surrounding sandstone and through connected OFC clusters. These effects balance even when no OFC cluster spans the reservoir boundaries. This conclusion is confirmed both for cases when the pressure gradient is along the $y$ (paleoflow) direction and for cases when it is along the $x$ direction.

The hierarchical and multiscale stratal architecture causes the connectivity of higher-permeability OFC cells to differ with scale and direction, and creates anisotropy in the bulk effective permeability. Since the difference in permeability between OFC and sandstone is up to 3 orders of magnitude and since the size of OFC clusters considerably exceeds the cell size, it could be expected that the water-oil boundary would show large-scale fingering. Although not included here, our results show that most of the oil reaches the production well through OFC clusters. Even so, we do not observe large-scale fingering (see Figures 7 and 8). In this type of stratal architecture, OFC cross strata are well distributed through unit bar and compound bar deposits, allowing oil to diffuse through them relatively uniformly.

Comparing Figures 7 and 8 shows that aligning the pressure gradient in the $x$-direction caused a broader and less-tapered waterflood front than aligning it in the $y$-direction. Thus, while the hierarchical organization of OFC cross strata within larger-scale deposits does not promote large-scale fingering, it does impart significant anisotropy. We found that the volume of water injected before water breakthrough is considerably larger (by a factor 1.5) when the pressure gradient is in the $x$ direction than when the gradient is in the $y$ direction, regardless of the pressure difference. When the pressure gradient is normal to paleoflow direction, more oil is recovered before water breakthrough (Figure 10, upper two lines in each graph) and also after injection of one movable pore volume of water (Figure 10, lower two lines in each graph). (Note that for the stratal architecture and reservoir geometry considered here, water production is greater than oil production by a factor ranging from 4 to 5 after injection of one movable pore volume of water.) This finding is consistent with general rule of waterflooding oil production, i.e. for enhanced sweep efficiency the line between injection and production wells should be normal to the line of directional permeability (Rose et al, 1989).

Although not presented here, our results show that neglecting capillary pressure during a simulation causes considerable error in both "local" state variables, such as oil saturation in an individual cell, and integral state variables such as oil production rate, water breakthrough time and oil sweep efficiency. As expected, the effective permeability is less when capillary pressure is represented because the pressure gradient must overcome the capillary pressure at oil-water boundaries.

The following are the main conclusions from this article:

1) Effective reservoir permeability and fluid mobility are essentially anisotropic because of the way open-framework conglomerates are organized within the multi-scale and hierarchical architecture in channel-belt deposits. As a result, total oil production, water breakthrough time, and oil sweep efficiency differ by a factor 1.5 for the different (horizontal) directions of pressure gradient. This result suggests that injection and production wells should be placed perpendicular to paleoflow flow direction in channel-belt deposits.

2) The value of the pressure gradient does not practically affect oil sweep efficiency although the spatial distribution of oil remaining in the reservoir depends on this value.

3) Oil sweep efficiency increases slightly as the proportion of open-framework conglomerate in the reservoir increases. When the proportion of OFC is lower than about 20%, clusters of connected OFC cells do not span the reservoir. Even so, the absence of spanning clusters of OFC does not abruptly affect oil sweep efficiency.

**ACKNOWLEDGMENTS**

We thank Schlumberger Limited for the donation of ECLIPSE Reservoir Simulation Software and the Ohio Supercomputer Center for technical support. This research was supported by the National Science Foundation under grant EAR-0810151. Any opinions, findings and conclusions or recommendations expressed in this article are those of the authors and do not necessarily reflect those of the National Science Foundation.

Hunt, A., & Ewing, R. 2009. *Percolation theory for flow in porous media*. Springer.

Huang, L., Ritzi, R.W. & Ramanathan, R. 2012. *Conservative models: Parametric entropy vs. temporal entropy in outcomes*. Ground Water .**50**, 199-206.

Jones, H. P., & Speers, R. G. 1976. *Permo-Triassic reservoirs of Prudhoe Bay field*, North Slope, Alaska.

Jamison, H.C., Brockett, L.D., McIntosh, R.A. 1980. *Prudhoe Bay-A ten year perspective*. In Halbouty, M.T., ed. *Giant oil and gas fields of the decade*. AAPG Memoir, **30**, 289-310.

Jones, A., Doyle, J., Jacobsen, T. & Kjonsvik, D. 1995, *Which sub-seismic heterogeneities influence waterflood performance? A case study of a low net-to-gross fluvial reservoir*. In De Haan, H.J., ed., *New Developments in Improved Oil Recovery*. London, Geological Society Special Publication, **84**, 5-18.

Jensen ,K.H., & Falta, R.W. 2005. *Chapter 2: Fundamentals*. In Mayer, A., and Hassanizadeh, S.M., eds. *Contamination of Soil and Groundwater by Nonaqueous Phase Liquids– Principles and Observations*. American Geophysical Union, Water Resources Monograph Series, **17**, 5-46.

Kortekaas, T.F.M. 1985. *Water/Oil displacement characteristics in crossbedded reservoir zones*. SPE Journal, **25,** 917 - 926.

Khataniar, S. & Peters, E. J. 1992. *The effect of reservoir heterogeneity on the performance of unstable displacements*. Journal of Petroleum Science and Engineering, **7**, 263-281.

Kjonsvik, D., Doyle, J., Jacobsen, T. & A. Jones. 1994. *The effects of sedimentary heterogeneities on production from a shallow marine reservoir - what really matters?* SPE Annual Technical Conference, 27-40.

Klingbeil, R., Kleineidam, S., Asprion, U., Aigner, T .& Teutsch, G. 1999. *Relating lithofacies to hydrofacies: outcrop-based hydrogeological characterization of quaternary gravel deposits*. Sedimentary Geology. **129**, 299– 310.

Kaasschieter, E.F. 1999. *Solving the Buckley-Leverett equation with gravity in a heterogeneous porous medium*. Computational Geosciences, **3**, 23-48.

Lunt, I. A. & Bridge, J. S. 2004. *Evolution and deposits of a gravelly braid bar and a channel fill, Sagavanirktok river, Alaska*. Sedimentology, **51**, 415– 432.
25